\documentclass[aps,prl,twocolumn,floats,superscriptaddress,showpacs]{revtex4}

\usepackage{graphicx}
\usepackage{amsmath,amssymb}
\usepackage{bm}

\begin{document}

\newcommand{\note}[1]{ { #1}}\newcommand{\nc}{\newcommand}\nc{\eq}{equation (\ref}
\nc{\h}{\hat} \nc{\hT}{\h{T}}\nc{\hV}{\hat{V}}\newcommand{\hH}{\hat{H}}\nc{\be}{\begin{eqnarray}}
\nc{\ee}{\end{eqnarray}}
\def\e{{\mathrm e}}
\def\ii{{\mathrm i}}
\def\dd{{\mathrm d}}
\def\lr#1#2#3{\left#1{#3}\right#2}
\def\bp{{\bm {p}}}
\def\br{{\bm {r}}}
\def\ef{{\varepsilon_{\text{\sc F}}  }}
\def\Tr{\operatorname{Tr}}

\title{Decoherence of charge qubit coupled to interacting background charges}

\author{Igor V.\ Yurkevich}
\affiliation{School of Physics and Astronomy, University of Birmingham,
Birmingham B15 2TT, United Kingdom}\affiliation{
Institut fur Nanotechnologie, Forschungszentrum Karlsruhe, 76021
Karlsruhe, Germany }

\author{Jim\ Baldwin}
\affiliation{School of Physics and Astronomy, University of Birmingham,
Birmingham B15 2TT, United Kingdom}

\author{Igor V.\ Lerner}
\affiliation{School of Physics and Astronomy, University of Birmingham,
Birmingham B15 2TT, United Kingdom}
\affiliation{Institute for Nuclear Theory,
University of Washington,
Seattle, WA 98195,
USA}

\author{Boris L. Altshuler}
 \affiliation{Physics Department, Columbia University,
538 West 120th Street, New York, N.Y. 10027, USA}
\affiliation{Institute for Nuclear Theory,
University of Washington,
Seattle, WA 98195,
USA}


\begin{abstract}{
The major contribution to decoherence of a double quantum dot or a Josephson junction charge qubit comes from the electrostatic coupling to fluctuating background charges hybridized with the conduction electrons in the reservoir.  However, estimations according to previously developed theories show  that finding a sufficient number of effective fluctuators in a realistic experimental layout is quite improbable. We show that this paradox is resolved by allowing for a short-range Coulomb interaction of the fluctuators with the electrons in the reservoir. This dramatically enhances both  the number of effective fluctuators and their contribution to decoherence, resulting in the most dangerous decoherence mechanism for charge qubits.}
 \end{abstract}

\maketitle

An implementation of quantum logical gates using solid-state nano-devices looks rather desirable due to their potential scalability. One of possible routes in this direction is developing of charge qubits. Schematically, a  charge qubit is a contact of two normal or superconducting islands with  charge carriers tunneling between them. The gate-controlled Coulomb blockade separates two charge states from all the others thus making a qubit \cite{shnirman,PlastinaFalci:03,AverinBruder:05}. Experimentally such charge qubits have been built as double quantum dots (DQD) \cite{hayashi,marcus,DQD:05,Andresen:07} or Josephson junction structures, made of a small superconducting island connected via a Josephson junction to  superconducting reservoirs \cite{naka,vion,yu,pash,Delsing:04,Astafiev}.

Decoherence  due to coupling to the environment remains a major impediment for developing charge qubits. This determines the demand for a better theoretical understanding of the  microscopic mechanisms that lead to decoherence. It is widely believed that the main contribution to decoherence in  charge  qubits comes from its coupling to fluctuating background  charges (FBCs) that create dynamical electric field, which affects qubit charge states (see ref.~\cite{Pashkin-review} for reviews of experimental evidence).  Since charge impurities are spatially quenched in the experimental temperature range $T\sim20\div50$mK,  the most probable cause for FBCs is from random recharging of electron traps embedded in an insulating layer close to the electronic bath.  The role of the bath can be played by  normal as well as superconducting metallic leads or gates when  the  electron states on the traps are hybridized with those in the bath. The appropriate model for FBCs has been developed \cite{paladino,galperin,faoro,GYL:05,vonDelft:05,galperin8,Montangero,paladino3,Marquardt} by analogy with the spin-fluctuator model of the spectral diffusion in glasses \cite{anderson}. In its frame, relevant experimental findings on the decoherence and dephasing in charge qubits have been successfully explained \cite{YBLA-1}.

The question which has not yet been addressed  is whether there is enough of  FBCs to cause experimentally observed decoherence. For a typical DQD qubit, the area of the electrodes is of order of $\mu m ^2$ and the volume available for the FBCs which can contribute to decoherence can be estimated as $10^{-15}\div10^{-16}{\text{cm}}^3$  so that their number is $N_{\text{geom}}\sim(10\div100)c$ where $c$ is their number concentration measured in ppm (parts per million).

As the appropriate electronic  levels are randomly spread in the energy interval of the order of the bandwidth, $D\sim1\div10$eV, the  effective number of traps $N_{\text{eff}}$ is reduced in comparison to $N_{\text{geom}}$  by a factor of $\delta/D$ where $\delta$ is the interval of energies in which FBCs are effective. Classical considerations \cite{galperin,paladino,faoro} suggest that $\delta$ is of the order of the relevant temperature, $T\sim 10\div50\mu$K, while the electronic trap levels separated from the Fermi sea by energies $\varepsilon_0\gtrsim T$ are frozen out, i.e.\ their contribution is exponentially suppressed.  Had this been the case,  $N_{\text{eff}}$ would be five orders in magnitude below $N_{\text{geom}}$. Then the probability to find even a single effective FBC would be negligible for any realistic value of $c$.

The full quantum considerations \cite{GYL:05,Marquardt} show  that  electron traps  energy levels up to $\varepsilon _0\lesssim\gamma_0$, contribute to decoherence. Here $\gamma_0$ is the hybridization energy, i.e.\ the broadening of the appropriate level. Its typical value can be much higher than $T$ but still is likely to be much smaller than $D$. The effective number of traps is thus
\begin{align}\label{nonwin}
    N_{\text{eff}} \sim\frac{\gamma_0}{D} N_{\text{geom}}\,.
\end{align}
Even with this enhancement $N_{\text{eff}} $ still seems to be of order $1$ or even smaller for cleaner samples, with $c\sim 1$ppm. At the same time experiments [8-13] show that even the cleanest samples suffer from considerable decoherence. Moreover,   assuming that FBCs, indeed, limit the decoherence and relaxation, one can  explain \cite{galperin} experimentally observed \cite{Astafiev} $1/f$ noise only \textit{provided that}  the number of FBCs is large. Thus it looks that the above considerations underestimate the efficiency of this source of decoherence.

In this letter we  show that allowing for the Coulomb interaction between the trapped charge and bath electrons  results in a dramatic increase of the effective hybridization rate $\gamma$ and thus in the upward revision of the estimate~(\ref{nonwin}).  We will present full analytic results that show how the interaction enhances decoherence.

We consider the model \cite{paladino,galperin} of the charge qubit indirectly coupled to the thermal bath  via the  coupling $\hat{V}=\sum_{i,\sigma} v_i\hat{d}^{\dagger}_{i\sigma}\hat{d}_{i\sigma}^{\phantom{{\dagger}} }$ to FBCs hybridized with the bath:
\begin{align}\label{HQ}
  \hat{H}_{Q}&= \big(\frac{\omega_0}{2}-\hat V\big)\hat{\tau}_z-\frac{E }{2}\hat\tau_x +\hat H_{\text{B}} .
\end{align}
Here $\omega_0=E_c(N-N_g)$ is the energy gap between the qubit  levels, $\hat\tau_i$ are the Pauli matrices in the space of qubit states, $E$ is the control energy (that includes the Josephson  energy in case of the JCQ); $\hat{d}^{\dagger}_{i\sigma}\,,\,\hat{d}_{i\sigma}^{\phantom{{\dagger}} }$ are the creation and annihilation operators for an electron with spin $\sigma=\uparrow,\downarrow$ at the $i^\text{th}$ trap and $v_i$ is   the electrostatic coupling of the qubit to the trapped electron.

We assume that the bath, described by $H_{\text{B}} $, is a normal metal -- the situation which is more relevant for the DQD charge qubit (for the JCQ this mechanism might be relevant as well since in the absence of screening  spatially remote trapped electrons hybridized with metallic gates  may contribute to decoherence). In this  case
\begin{align}\label{HB}
    \hat H_{\text{B}}  = \hat H_{\textrm{met}}+ \hat H_{\textrm{imp}}+ \hat H_{\text{int}}\,.
\end{align}
Here $\hat H_{\textrm{met}}=\sum_{\bp,\sigma}^{}\varepsilon _{\bp}\hat c^{\dagger}_{\bp,\sigma} \hat c^{\phantom{\dagger} }_{\bp,\sigma}$ is the   Hamiltonian of  the metallic bath, $\hat c^{\dagger}_{\bp,\sigma}\,,\; \hat c^{\phantom{\dagger} }_{\bp,\sigma}$  creation and annihilation operators for bath electrons. The energy $\varepsilon _\bp\equiv p^2/2m-\ef$ is counted from the chemical potential (and so are $\varepsilon _i$ below).  The impurity Hamiltonian describes random levels  which can trap electrons,
\begin{eqnarray}\label{imp}
 \hat H_{\textrm{imp}}= \sum_{i  }\left [ \varepsilon_i\, \hat n_{i}+
  U^H_i \hat n_{i\uparrow} \hat n_{i\downarrow} \right ]\,,
\end{eqnarray}
where $ \hat n_{i\sigma} = \hat{d}^{\dagger}_{i\sigma}\hat{d}^{\phantom{\dagger} }_{i\sigma}$ and $\hat{n}_i=\hat{n}_{i\uparrow} + \hat{n}_{i\downarrow}$. The Hubbard on-site repulsion $U^H_i$ is assumed to be very large so that two electrons cannot reside on the same site. The relevant charge fluctuations are due to switchings  between empty and single-occupied impurity states. In what follows we disregard double-occupied states and omit irrelevant spin indices, absorbing the spin degeneracy of the bulk electrons into the density of states $\nu_0\equiv1/\mathcal{V}\delta  $.

The $d$-electrons are coupled to the bath  via direct hybridization as well as via the local Coulomb interaction:
\begin{eqnarray}\label{int}
\hat H_{\text{int}}= \sum_{i, \bp }^{}\left [ t_{i }^{\phantom{{\dagger}}} \hat c^{\dagger}_{\bp }\hat d^{\phantom{{\dagger}} }_{i }   +\text{h.c.}   \right ] + \sum_{i}^{} U _i\hat{n}_i\hat{\rho}_i\,.
\end{eqnarray}
Here $t_i$ is the tunneling amplitude from the  resonant level at the $i^{\text{th}}$ trap into the bulk, $\hat{\rho}_i$ is the bath  electron density operator at the $i^{\text{th}}$ site and the interaction $U_i$ is positive for an acceptor and negative for a donor  level.

The model of Eqs.~(\ref{HQ})-(\ref{int}) was previously considered \cite{galperin,paladino,faoro,GYL:05,vonDelft:05,%
galperin8,Montangero,paladino3,Marquardt} without the interactions $U_i$ in Eq.~(\ref{int}). We will   extend the non-perturbative procedure developed in \cite{GYL:05} to evaluate the contribution to decoherence of the charge qubit due to FBCs interacting with the bath electrons.

Let the full density matrix of the system initially be separable, i.e.\ $\hat{\rho}_{\text{B+Q}}(0)=\hat{\rho}
(0)\otimes\hat{\rho}_{\text{B}} $, where $\hat\rho(t)$ is the
reduced density matrix of the qubit and
$\hat\rho_{\text{B}}=Z_{\text{B}}^{-1}\e^{-{\beta} \hat
H_{\text{B}}}$ is the equilibrium density matrix of the bath. Then one can formally represent $\hat{\rho}
(t)$ as
\begin{align}
\hat{\rho} (t)
= \begin{pmatrix}  n(t) & \rho _{12}(0)\e^{-i\omega_0t}\mathcal{D}(t)\\
\rho _{21}(0)\e^{ i\omega_0t}\mathcal{D}^*(t)(t) & 1-n(t)\end{pmatrix}.
\label{reduced}
\end{align}
In the long-time limit, both diagonal and off-diagonal elements of $\hat{\rho}
(t)$ get exponentially close to their equilibrium values. The corresponding decay rates are usually referred to as the relaxation, $\Gamma_1$, and decoherence, $\Gamma_2$, rates.
Below we discuss ``pure dephasing", i.e.\ let $E=0$ in Eq.~(\ref{HQ}). This is   justified for the charge qubit during the most of the operational cycle \cite{paladino}.  We find
the time evolution of $\mathcal{D}(t)$ as
\begin{align}\label{f}
\mathcal{D}(t)=\left\langle \e^{i(\hat{H}_B+\hat{V})t
}\,\e^{-i(\hat{H}_B-\hat{V})t }\right\rangle
\end{align}
where $\langle\hat{A}\rangle \equiv \Tr({\rho_B \hat{A}})$. As $\Gamma_2=-\lim_{t \to\infty}\,\,t ^{-1}\ln\left|D(t
)\right|$, we need to calculate  the asymptotical value of $\mathcal{D}(t)$ at $t\to\infty$. To this end, we represent $\mathcal{D}(t)$ in Eq.~(\ref{f}) as a functional integral over the Keldysh double-time contour, which is natural as Eq.~(\ref{f}) contains time-ordered and  anti-time-ordered exponentials.

Note now that Eqs.\ (\ref{HB})--(\ref{int}) determine a well known problem of the interacting resonant level   (IRL)  which can be mapped \cite{toulouse} onto the single channel Kondo problem. This suggests that one can account for the interaction (the second term in Eq.~(\ref{int})) by using the renormalization group approach. We will show that this leads only to substituting the renormalized (energy-dependent) hybridization rate for the bare one into Eq.~(\ref{finiteT}) below.

In the absence of the interaction term in Eq.~(\ref{int}) one performs an exact functional integration over the fields describing electrons in the bath. This results in the effective action in terms of  the fields $\bar d_{i }(t),\; d_{i }(t)$ which describe trapped electrons
\begin{align}\label{S}
     \mathcal{S}[d^{\dagger},d]&=\sum_{ij}^{}\int\!\!\dd t\,\dd t'\,d^{\dagger}_i(t)\hat{G}^{-1}_{ij}(t,t')d_j(t')\,.
\end{align}
The time integrations are performed along the standard Keldysh contours. The Keldysh Green function $\hat{G}_{ij}(t,t')$ of the FBC is defined by  the  Dyson equation resulted from the integration over the bath electron fields:
 \begin{align}\notag
     \hat G^{-1}_{ij}(t,t')&=\delta(t,t')\delta_{ij}[{\ii \partial_t \!-\! \varepsilon _i \!     +\! v_i(t,t')}]-\hat {\Sigma}_{ij}(t,t'),\\ \label{D}
\hat {\Sigma}_{ij}(t,t')&=t_it_j^* \hat{g}({\br}_i-{\br}_j'; t-t')\,.
 \end{align}
Here $v_i(t,t')=\pm \theta(t)\theta(t-t')v_i$ with `$+$' sign on the upper and `$-$' sign on the lower branch of the Keldysh contour, and  $\hat{g}$ is the bare Keldysh Green function of the conduction  electrons: the Fourier transforms of its retarded component is $g^R(\varepsilon,\bp)=(\varepsilon-\varepsilon _\bp +\ii 0)^{-1}$.  Typically,  FBCs are separated by a distance far exceeding $\lambda_{\text{F}}$. Then the off-diagonal contributions to $\Sigma_{ij}$ is suppressed upon a spatial integration. Thus $\hat G_{ij}=\hat G_i\delta_{ij}$ with $G^R_i(\varepsilon)= ({\varepsilon-\varepsilon _i + \ii \gamma_i/2})^{-1}$ and $G^K_i(\varepsilon) = 2\operatorname{Im}G_i^R(\varepsilon)\tanh(\varepsilon/2T) $. The width  $\gamma_i=\operatorname{Re}\Sigma_{ii} =\pi  |t_i|^2 /\delta$  of the  $i^{\text{th}}$ trap level   plays the role of the recharging switching rate.

Since contributions of different FBCs are uncorrelated, we consider a single fluctuator with the switching rate $\gamma_0$. Then the functional averaging in Eq.~(\ref{f}) with the action (\ref{S})  results in $$
  \Gamma_2 =-\textrm{Re}\lim_{t\rightarrow\infty}t^{-1} \Tr\ln (1+ \hat{v}\hat{G} ).$$
After expanding this in powers of $\hat{v}\hat{G}$ , one finds that the $t^{-1}$ factor cancels while the remaining time-integrals in each order have the convolution structure \cite{GYL:05}. This allows to re-exponentiate  the expansion after performing the Fourier transform.  Taking then the trace in the Keldysh space, we arrive  at
\begin{align}
\Gamma_2=-\Re \!\int\limits_{-\infty}^{+\infty}\frac{\dd\varepsilon}{2\pi}
\ln \left[1+{v} G^{\text K}(\varepsilon)-{v}{G}^{\text R}(\varepsilon){v}{G}^{\text
A}(\varepsilon)\right].\label{Det}
\end{align}
Substituting   $ {G}^{R,A}$ and $G^{K}$ into Eq.~(\ref{Det}) we find $\Gamma_2$ as a function of $T$ and $\gamma(\varepsilon)$:
\begin{subequations}\label{fT}
\begin{equation}\label{finiteT}
  \Gamma_{2} =-\!\!\int\limits_{-\infty}^{\infty}\frac{\rm{d}\varepsilon}{4\pi}
  \ln\left\{1-4f({\varepsilon })[1-f({\varepsilon })]{\sin^{2}\varphi (\varepsilon)}\right\},
\end{equation}
where $f(\varepsilon) \equiv (1+\e ^{\varepsilon /T})^{-1}$ is the Fermi factor and
\begin{equation}\label{finiteT1}
  \varphi (\varepsilon)= \arctan \frac{2v\gamma(\varepsilon)}{(\varepsilon-\epsilon_{0})^{2}+ \gamma^{2}(\varepsilon)-v^{2}}\;.
\end{equation}
\end{subequations}

Note that the interaction in Eq.~(\ref{int}) entered $\Gamma_2$ only through substituting the renormalized energy-dependent switching rate $\gamma(\varepsilon )$ for the bare value $\gamma_0$. The RG equations for the IRL model, Eqs.~(\ref{HB})--(\ref{int}), have been obtained by mapping the tunneling amplitude $t_0$ onto the transverse coupling $J_{\perp}$ in the Kondo model (see, e.g., \cite{Schlottmann,matveev,zawad}). These RG equations are equivalent to those originally derived for the Kondo model within the 1D Coulomb gas picture \cite{YA}. It is convenient to write them in terms of the effective charges $Y$ and $\delta$,
 \begin{align*}
\frac{{\rm d}Y}{{\rm d}\ln \epsilon}&=\left[-1-2\frac{\delta}{\pi}+\left(\frac{\delta}{\pi}\right)^2\right]Y, &  \frac{{\rm d}\delta}{{\rm d}\ln \epsilon}&=\left(1-\frac{\delta}{\pi}\right)Y^2,
\end{align*}
where  $\delta\equiv2\arctan\left(\pi\nu_0U/2\right)$, $
Y\equiv(\gamma/\epsilon)\cos^2\delta
$ and $\epsilon$ is the running energy cutoff. The upper cutoff is the bandwidth $D$ of the bath electrons. As  typically  $\gamma_0 \ll D$, the bare value $Y_0\sim \gamma_0/D\ll1$ so that one can neglect the renormalization of $\delta$. Substituting  $\delta=\delta_0$, we rewrite the RG equations as
\begin{align}\label{rg}
    \frac{{\rm d}\ln\gamma}{{\rm d}\ln \epsilon}&=-\alpha, & \alpha&\equiv2\,\frac{\delta_0}{\pi}-\left(\frac{\delta_0}{\pi}\right)^2\,.
\end{align}
Equation (\ref{rg}) yields
 $\gamma(\varepsilon)=\gamma_0( {D}/{\varepsilon})^{\alpha}.$
The energy dependence saturates for small $\varepsilon$ at $\varepsilon=\max (\gamma,T,\varepsilon_0)$. The parameter $\alpha$ can be positive or negative, depending on the interaction sign. For $U_0>0$ the linear in $\delta$ contribution to $\alpha$ (due to the Mahan exciton) is dominant \cite{Mahan} and exceeds the quadratic contribution (due to the Anderson orthogonality catastrophe) so that $\alpha>0$. Below we take into account only the levels with positive $\alpha$.

According to Eq.~(\ref{fT}) all traps with $\gamma(\varepsilon )>\varepsilon _0$   contribute effectively to $\Gamma_2$. Since   $\gamma({\varepsilon })\gg \gamma_0$ for $\alpha>0$, the fraction of effective FBCs dramatically increases in comparison to the non-interacting case.

At a given $T$ each FBC is either in the high-temperature, $T\gtrsim \gamma(\varepsilon )$, or in the low-temperature, $T \lesssim \gamma({\varepsilon })$, regime, depending on the relation between $T$ and
 $
 T_0\equiv \gamma_0\left(  D/{\gamma_0 }\right)^{\frac\alpha{1+\alpha}}
 $. As $\varepsilon _0\lesssim \gamma_0$ the effective number, Eq.~(\ref{nonwin}), of ``high-T'' FBCs can be estimated as  $ (T/D) ^{1+\alpha} N_{\text{geom}}$, making their presence  rather improbable -- but not impossible. The switching rate for such FBCs  is $T$-dependent and saturated at
$
    \gamma(T)=\gamma_0\left(  D /T\right)^{\alpha}
$.
Most (if not all) FBCs are in the low-$T$ regime, $T\lesssim \gamma(\varepsilon )$, i.e.\  characterized by $ T_0$ exceeding $T$. In this case the switching rate  $\gamma(\varepsilon )$  saturates at   $
    \gamma =\gamma_0\left(D/\gamma\right)^{\alpha/(1+\alpha)} \,.
$

\begin{figure}[t]
\includegraphics[width=0.9\columnwidth]{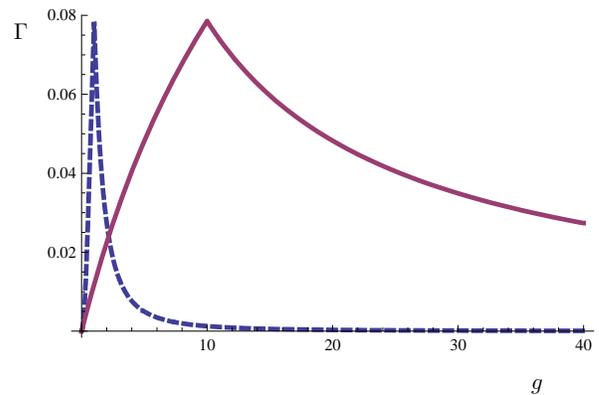}
\caption{Decoherence rate as a function of coupling strength, $g\equiv v/\gamma$, for $T=0.1\gamma$, $D=10\gamma$.  The dashed line shows the non-interacting case ($\alpha=0$), while the solid line describes decoherence with allowance for the interaction ($\alpha=0.5$) which turns out to be substantial in a much wider parametric range.}
\end{figure}

Since at low $T$ the product of Fermi factors in Eq.~(\ref{finiteT}) is a peak around zero (i.e.\ the Fermi energy) of  width $T$, we  approximate $\gamma({\varepsilon })$ by the $\varepsilon $-independent expression,
\begin{align}\label{gamma}
    \gamma\approx \gamma({\varepsilon=\gamma_0 })= \left\{\begin{array}{ll}
                              \gamma_0 \left(D/\gamma_0\right)^{\frac{\alpha}{\alpha+1}},& {\gamma\gtrsim\varepsilon_{0}} \\
                                \gamma_0 \left(D/\varepsilon_{0}\right)^{\alpha}, & {\gamma\lesssim\varepsilon_{0}}
                             \end{array}\right.
                             \end{align}
Thus the contribution of the low-$T$ FBC into the decoherence rate is
\be\label{Ga2}
  \Gamma_{2}^\text{low} (T)=\frac{T}{\pi}\arctan^{2}\left[\frac{2v\gamma} {\varepsilon_{0}^{2}+\gamma^{2}-v^{2}}\right].
\ee
Figure~1 illustrates this behavior and shows that the interaction dramatically increases the energy where FBCs effectively contribute to $\Gamma_2$.
For $\varepsilon_{0}\gg\gamma$ the decoherence rate decays with $\varepsilon _0$  faster than in the non-interacting case: $\Gamma_2\propto\varepsilon_{0}^{-4-2\alpha}$. This means that, as was noticed above,  only  FBCs with   $\varepsilon_{0}\lesssim\gamma$ cause decoherence. Using $\gamma$ from the upper line in Eq.~(\ref{gamma}) we find  the effective number of FBCs:
\begin{align}\label{win}
    N_{\text{eff}} \sim\left(\frac{\gamma_0}{D}\right)^{\frac1{1+\alpha}} N_{\text{geom}}\,.
\end{align}
This is considerably higher than that in the absence of the interaction, Eq.~(\ref{nonwin}).

We have already noticed that it is statistically improbable to find the FBCs in the high-$T$ regime. Had they  existed,  their contribution  would be also  enhanced by the interaction  as compared to the expressions obtained classically \cite{paladino,galperin}, namely
 \begin{equation}
\Gamma_2^{\text{high} }(T)=
\gamma(T)-\theta(\gamma(T)-v)\sqrt{\gamma^2(T)-v^2}\, ,\label{noT}
\end{equation}
where $\gamma(T)=\gamma_0\left(  D /T\right)^{\alpha}$.
Eq.~(\ref{noT})  predicts a non-analytic in $T$ decoherence rate with a well pronounced peak  at $T=Dg^{-1/\alpha}$.

In conclusion, we have demonstrated that the fluctuating background charges -- namely, electrons trapped in the insulating layer close to the surface of metallic gates/electrodes and thus hybridized and interacting with the conduction electrons -- cause significant decoherence of the charge qubit. The interaction leads to a drastic increase of the efficiency of this mechanism of decoherence and to a parametric enhancement of the resulting contribution, as well as of the number of the effective FBCs given by Eq.~(\ref{win}).  This enhancement proves the assumption that the FBCs act as the most dangerous source of decoherence for the charge qubit.

\begin{acknowledgments}
We thank Yuri Galperin for useful discussions. This work
was supported by the EPSRC grant  T23725/01 and by the US DOE contract No.\ DE-AC02-06CH11357.
\end{acknowledgments}
%


\begin{thebibliography}{10}
\bibitem{YBLA-1}
The most probable alternative microscopic mechanism of decoherence, via the
  qubit coupling to the phonon bath, is not  efficient for the
  charge qubit, provided that one chooses a realistic phonon spectra
  \cite{baranger}.
\end{thebibliography}

\end{document}